\documentclass[aps, prd, reprint, amsfonts, amssymb, amsmath, showkeys, nofootinbib, twoside, superscriptaddress, longbibliography]{revtex4-1}

\usepackage{caption, subcaption}
\usepackage[pdftex, pdftitle={Article}, pdfauthor={Author}]{hyperref} 
\usepackage{amsthm, amssymb}
\usepackage{mathtools}
\usepackage{acronym}
\usepackage{xcolor}
\usepackage{graphicx}
\usepackage{xspace}
\usepackage{array}
\usepackage{float}

\graphicspath{figures}

\usepackage{multirow}
\usepackage{array, boldline}
\newcolumntype{P}[1]{>{\centering\arraybackslash}p{#1}}
\usepackage[verbose]{placeins}

\newcommand{\ssb}[1]{{\color{cyan} SB: #1}}

\usepackage{array}
\newcolumntype{P}[1]{>{\centering\arraybackslash}m{#1}}
\renewcommand*\arraystretch{1.3}

\begin{document}

\title{The unresolved stochastic background from compact binary mergers detectable by next-generation ground-based gravitational-wave observatories}


\author{Darsan S. Bellie}
\email[Email: ]{darsh@northwestern.edu}
\affiliation{Center for Interdisciplinary Exploration and Research in Astrophysics (CIERA), Northwestern University, 1800 Sherman Ave, Evanston, IL 60201, USA }
\affiliation{Department of Physics and Astronomy, Northwestern University, 2145 Sheridan Road, Evanston, IL 60201, USA}

\author{Sharan Banagiri}
\affiliation{Center for Interdisciplinary Exploration and Research in Astrophysics (CIERA), Northwestern University, 1800 Sherman Ave, Evanston, IL 60201, USA }

\author{Zoheyr Doctor}
\affiliation{Center for Interdisciplinary Exploration and Research in Astrophysics (CIERA), Northwestern University, 1800 Sherman Ave, Evanston, IL 60201, USA }

\author{Vicky Kalogera}
\affiliation{Center for Interdisciplinary Exploration and Research in Astrophysics (CIERA), Northwestern University, 1800 Sherman Ave, Evanston, IL 60201, USA }
\affiliation{Department of Physics and Astronomy, Northwestern University, 2145 Sheridan Road, Evanston, IL 60201, USA}

\date{\today}

\begin{abstract}

The next generation of ground-based gravitational-wave detectors will look much deeper into the Universe and have unprecedented sensitivities and low-frequency capabilities. Especially alluring is the possibility of detecting an early-Universe cosmological stochastic background that could provide important insights into the beginnings of our Universe and fundamental physics at extremely high energies. However, even if next-generation detectors are sensitive to cosmological stochastic backgrounds, they will be masked by more dominant astrophysical backgrounds, namely the residual background from the imperfect subtraction of resolvable compact binary coalescences (CBCs) as well as the CBC background from individually unresolvable CBCs. Using our latest knowledge of masses, rates, and delay time distributions, we present a data-driven estimate of the unresolvable CBC background that will be seen by next-generation detectors. Accounting for statistical and systematic errors, this estimate quantifies an important piece in the CBC noise budget for next-generation detectors and can help inform detector design and subtraction algorithms. We compare our results with predictions for backgrounds from several cosmological sources in the literature, finding that the unresolvable background will likely be a significant impediment for many models. This motivates the need for simultaneous inference methods or other statistical techniques to detect early-Universe cosmological backgrounds. 

\end{abstract}

\maketitle

\acrodef{gw}[GW]{gravitational wave}
\acrodef{sgwb}[SGWB]{stochastic gravitational wave background}
\acrodef{ce}[CE]{Cosmic Explorer}
\acrodef{et}[ET]{Einstein Telescope}

\section{Introduction} 
\label{sec:intro}



In 2015, the Laser Interferometer Gravitational-Wave Observatory (LIGO)~\cite{LIGOScientific:2014pky} directly detected \acp{gw} for the first time by capturing the coalescence of two stellar-mass binary black holes (BBHs)~\cite{LIGOScientific:2016aoc}. This was eventually followed by the detection of GWs from binary neutron stars (BNSs)~\cite{LIGOScientific:2017vwq} and neutron-star black-hole binaries (NSBHs)~\cite{LIGOScientific:2021qlt} by the LIGO and Virgo observatories~\cite{VIRGO:2014yos}. Constituting the category of compact binary coalescences (CBCs), these sources have heralded the arrival of \ac{gw} astronomy with nearly a hundred detected candidates already~\cite{LIGOScientific:2018mvr, LIGOScientific:2020ibl, LIGOScientific:2021djp}. 

While we have only detected GWs from CBCs so far, several other sources have been theorized and expected. In particular, several models postulate the presence of ``cosmological" \acp{gw}, predominantly showing up as a cosmological \ac{sgwb}~\cite{Caprini:2018mtu, Christensen:2018iqi, Sakellariadou:2022tcm}. These might be accessible by next-generation ground-based GW detectors (henceforth known as XG detectors) like \ac{ce}  and \ac{et}~\cite{Evans:2021gyd, Punturo:2010zz, Regimbau:2016ike, Maggiore:2019uih}. Cosmological \acp{sgwb} originating from the early Universe might carry imprints of the physics of the earliest epochs in the Universe, inaccessible through any other channels. Detection of such a background would therefore drastically expand our understanding of the first instants in the Universe and unlock fundamental physics at very high energies well beyond the reach of particle accelerators~\cite{Caprini:2018mtu}. 

However, such cosmological \acp{sgwb} can be shielded by the \acp{gw} arising from various astrophysical sources~\cite{Regimbau:2011rp, Regimbau:2008nj, Buonanno:2004tp} and in particular CBCs of stellar origin since they are expected to be the largest contributor to the \ac{gw} power in the frequency band of XG detectors. The \acp{gw} from CBCs will pose a major challenge for the capability of XG detectors to detect cosmological backgrounds. There are two components of such astrophysical shielding, coming from ``resolved" and ``unresolved" CBCs. Resolved sources, here, refer to those sources with \ac{gw} signals that can be individually and confidently detected within the detector noise. 

The proposed XG detectors are expected to have a high redshift reach to astrophysical CBC sources owing to their better sensitivities and low-frequency capabilities of up to $\sim$ 5 Hz~\cite{Kalogera:2019sui, Evans:2023euw, Evans:2021gyd, Kalogera:2019sui}. It is expected that XG detectors would be able to resolve nearly all BBHs of stellar origin in the Universe as well as many BNSs and NSBHs up to redshifts of a few~\cite{Kalogera:2019sui, Evans:2021gyd, Evans:2023euw}. To minimize their impact on searches for cosmological \acp{sgwb}, significant ongoing work is devoted to developing techniques to subtract resolvable CBCs from \ac{gw} data~\cite{Zhou:2022nmt, Zhou:2022otw, Zhong:2022ylh, Sachdev:2020bkk, Regimbau:2022mdu, Sharma:2020btq}. However, the subtraction residue from imperfect subtraction could still be a challenge for XG detectors~\cite{Zhou:2022otw, Zhou:2022nmt}. 

In addition to the resolvable CBC signals, we also expect an astrophysical  \ac{sgwb} originating from the incoherent superposition of unresolvable CBC sources. Such an astrophysical  \ac{sgwb} is a key detection target for current-generation ground-based detectors~\cite{LIGOScientific:2016jlg, LIGOScientific:2019vic, KAGRA:2021kbb}, but it can also shield cosmological \acp{sgwb}. Crucially, this unresolved CBC background represents a noise source that is independent of the fidelity and efficacy of subtraction. Accurately characterizing this background will enable us to understand the accessibility of cosmological \acp{sgwb} to XG detectors, will inform us of the necessary levels of subtraction of resolvable signals, and will provide an important data point in comparing detector designs and networks. Therefore, in this paper, we provide a data-driven estimate of the levels of the unresolvable CBC  \ac{sgwb} expected for different configurations of XG detector networks. We draw upon several models used in the population inference with LIGO-Virgo-KAGRA (LVK) observations and Galactic binary neutron stars; by utilizing multiple models, we provide robust estimates that account for both statistical and systematic uncertainty. 

The rest of the paper is organized as follows. In Sec.~\ref{sec:unresolved_CBCs} we briefly motivate the paper and place our study in the context of previous literature for XG detectors. In Sec.~\ref{sec:methods_context}, we mathematically define the \ac{sgwb} and describe how an astrophysical \ac{sgwb} can be calculated from a set of sources. Sec.~\ref{sec:pop_models} describes our assumptions on the astrophysical merger rates, mass distributions, and redshift distributions which we use to compute the unresolved background, and Sec.~\ref{sec:detector_networks} describes our choices of detector network configurations. In Sec.~\ref{sec:methods_2}, we detail our formalism to numerically compute the unresolved background from CBC sources and in Sec.~\ref{sec:cosmo_backgrounds}, we discuss the cosmological \ac{sgwb} models that we use. Sec.~\ref{sec:results} presents our results, and Sec.~\ref{sec:conclusion} discusses their implications and comments on future work needed.

Throughout this paper, $G$ is the gravitational constant, $c$ the speed of light, $H_0$ the Hubble constant, and we define our cosmology using the $\Lambda$CDM model with cosmological parameters taken from Planck 2018~\cite{Planck:2018vyg}.

\section{The unresolved SGWB from CBCs}
\label{sec:unresolved_CBCs}

The total GW energy-density spectrum $\Omega_{\text{GW}}$ as seen by a GW detector is the sum of the GW energy-density spectrum from cosmological sources $\Omega_{\text{cosmo}}$ and from astrophysical sources~\footnote{i.e. of stellar origin.} $\Omega_{\text{astro}}$~\cite{Sachdev:2020bkk}, 
\begin{equation}
\label{eqn: omega_breakdown_1}
\Omega_{\text{GW}} = \Omega_{\text{astro}} + \Omega_{\text{cosmo}}.
\end{equation}
In general, the astrophysical component can mask the cosmological component if $\Omega_{\text{astro}} > \Omega_{\text{cosmo}}$. For the rest of this paper, we assume that $\Omega_{\text{astro}} = \Omega_{\text{cbc}}$ and that the \ac{gw} energy from other sources can be neglected so that 
\begin{equation}
\label{eqn: omega_breakdown_2}
\Omega_{\text{GW}} \approx \Omega_{\text{cbc}} + \Omega_{\text{cosmo}}.
\end{equation}

Various methods have been developed to subtract resolvable CBC signals in order to minimize their impact on searches for $ \Omega_{\text{cosmo}}$~\cite{Zhou:2022nmt, Zhou:2022otw, Zhong:2022ylh, Sachdev:2020bkk, Regimbau:2022mdu, Sharma:2020btq}. Such techniques, however, are always imperfect, as they depend on parameter estimation results and can also suffer from waveform systematics. The imperfect subtraction leaves behind a residue $\Omega_{\text{cbc, residue}}$ that contributes to the effective \ac{sgwb} from CBCs~\cite{Sachdev:2020bkk, Zhou:2022nmt}.

In addition to the resolved CBC signals, the incoherent superposition of the unresolvable CBC signals gives rise to an astrophysical \ac{sgwb} that we call $\Omega_{\text{cbc, unres}}$. Therefore, effectively,
\begin{equation}
    \Omega_{\text{cbc}} = \Omega_{\text{cbc, unres}} + \Omega_{\text{cbc, residue}}. 
    \label{Eq:cbc_forground}
\end{equation}
While some studies find that $\Omega_{\text{cbc, residue}}$ remains a major challenge for detecting cosmological \acp{sgwb}~\cite{Zhou:2022nmt, Zhou:2022otw}, others find that $\Omega_{\text{cbc, unres}}$ will provide the noise floor~\cite{Sharma:2020btq, Zhong:2022ylh}. Since $\Omega_{\text{cbc, unres}}$ is a purely stochastic noise, it will have to be fit simultaneously with any cosmological \ac{sgwb}.

While some constraints were previously placed on $\Omega_{\text{cbc, unres}}$ for XG detectors by Refs.~\cite{Zhou:2022nmt, Sachdev:2020bkk, Regimbau:2016ike, Wu:2011ac, Zhong:2022ylh, Zhu:2012xw} they were either limited by the dearth of data before O3, or by simple population models or both. Ref.\cite{Zhou:2022otw} in particular accounts for the uncertainties in the latest LVK BBH and BNS rate inferences in their calculation of the residue but not for $\Omega_{\text{cbc, unres}}$. Furthermore, papers in the literature generally do not include the contribution from NSBHs, whose unresolved \ac{sgwb}, we find, is comparable with that of BNSs~\footnote{Although Ref.~\cite{Zhu:2012xw} includes NSBHs in their estimate, this was not data-informed.}.

In our study, we implement a data-driven estimate of $\Omega_{\text{cbc, unres}}$ for XG detectors, using the latest inferences on the CBC population from LVK data. Wherever possible, we consistently incorporate the uncertainty from the rate and from the population modeling. In addition, we consider several population models to propagate the systematic uncertainty stemming from parameterized GW analyses to present uncertainty envelopes for $\Omega_{\text{cbc, unres}}$. 

\section{Calculating the SGWB from CBCs}
\label{sec:methods_context}

 The  \ac{sgwb} from any type of GW source is generally characterized by the dimensionless GW energy-density spectrum~\cite{Romano:2016dpx}
\begin{equation}
\label{eqn: omega_gw_general}
\Omega_{\text{GW}} (f) = \frac{f}{\rho_c} \frac{d\rho_{\text{GW}}}{df} (f),
\end{equation}
where $f$ is the observed GW frequency, $\rho_c = 3H_0^2c^2/8\pi G$ is the critical density needed to close the Universe, and $\rho_{\text{GW}}$ is the GW energy density at $f$.

The GW energy density $\rho_{\text{GW}}$ arising from astrophysical sources is given by 
\begin{equation}
\label{eqn: rho_flux_basic}
\frac{d\rho_{\text{GW}}}{df} (f) = \frac{1}{c} F(f),
\end{equation}
where $F$ is the GW energy flux in the observer frame~\cite{Ferrari:1998jf}. We define the energy flux by summing up the individual fluxes in a population of astrophysical sources between redshifts $z_{\rm low}$ to $z_{\rm up}$ and with an associated set of source parameters $\theta$ as~\cite{Regimbau:2011rp, Regimbau:2022mdu} 
\begin{equation}
\label{eqn: omega_gw_flux_def}
F(f) = \int_{\theta} p(\theta) d\theta \int_{z_{\text{low}}}^{z_{\text{up}}} \frac{dR_o(z)}{dz} \frac{\frac{dE_{\text{gw}} (f_{\text{s}}, \theta)}{df_{\text{s}}} (1+z)^2}{4\pi d_L^2} dz,
\end{equation}
where $p(\theta)$ is the distribution of the source parameters and $d_L$ is the luminosity distance. The source-frame GW energy spectrum emitted by each astrophysical source is given by ${dE_{\text{gw}} (f_{\text{s}}, \theta)}/{df_{\text{s}}}$, where $f_{\text{s}} = f(1+z)$ is the source-frame frequency. 

For astrophysical sources distributed isotropically across the Universe, the rate of observed signals $R_o$ in some redshift slice $dz$ is
\begin{equation}
\label{eqn: observed_rate_converted}
    \begin{multlined}
    \frac{dR_o(z)}{dz} = \frac{dN(z)}{dz\:dt_o} = \frac{dN(z)}{dV_c\:dt_s} \frac{dt_s}{dt_o} \frac{dV_c}{dz} (z)\\
    = R_{\nu}(z) \frac{1}{1+z} \frac{dV_c}{dz} (z),
    \end{multlined}
\end{equation}
where $N$ specifies the number of events occurring in a given cosmic slice and $dt_s/dt_o = 1/(1+z)$ accounts for time dilation between the source and observer. $R_{\nu}(z) = dN(z)/dV_c\:dt_s$ is the source-frame rate density per comoving volume $V_c$ of a specific type of astrophysical source $\nu$. The differential comoving volume element is given by~\cite{Dominik:2014yma}
\begin{equation}
\label{eqn: comoving_volume_element}
\frac{dV_c}{dz} (z) = \frac{4\pi d_L^2}{(1+z)^2} \frac{c}{H_0 E(z)},
\end{equation}
where, for a flat $\Lambda$CDM cosmology and ignoring radiation density, $E(z) = \sqrt{\Omega_M (1+z)^3 + \Omega_\Lambda}$~\cite{Regimbau:2011rp, Regimbau:2022mdu}. Bringing this all together, we arrive at the energy-density spectrum for a  \ac{sgwb} arising from astrophysical sources distributed isotropically across the Universe between some redshifts $z_{\rm low}$ to $z_{\rm up}$:

\begin{equation}
\label{eqn: omega_gw_final}
\Omega_{\text{GW}} (f) = \frac{f}{\rho_c H_0} \int_{\theta} \int_{z_{\text{low}}}^{z_{\text{up}}} \frac{R_{\nu}(z)\frac{dE_{\text{gw}} (f_{\text{s}}, \theta)}{df_{\text{s}}} p(\theta)}{(1+z)E(z)} d\theta dz.
\end{equation}

Finally, the spectral energy $E_{\text{gw}}$ emitted by any astrophysical source can be related to the amplitudes of the plus ($+$) and cross ($\times$) GW polarizations $\tilde{h}_{+}$ and $\tilde{h}_{\times}$ via~\cite{Phinney:2001di}
\begin{equation}
\label{eqn: spectral_energy_flux_conversion}
\frac{d E_{\rm gw}}{df_s}= \frac{2 \pi^2 c^3 d^2_L f^2 }{G (1 + z)^2}  \biggr<|\tilde{h}_{+}(f, \theta)|^2 + |\tilde{h}_{\times}(f, \theta)|^2 \biggr>_{\Omega_s},
\end{equation}
where the right-hand side is averaged over the source orientations ${\Omega_s}$.

\section{The CBC population models}
\label{sec:pop_models}

In the previous sections, we provided the motivation and the methods to calculate $\Omega_{\text{cbc, unres}}$, the unresolved \ac{sgwb} from CBCs. Now we describe the models we use for the astrophysical population of CBCs, focusing in particular on detectability by XG GW detectors. Several studies have shown that almost all of the BBHs will be individually resolvable by these detectors~\cite{Evans:2021gyd, Evans:2023euw, Zhou:2022nmt} and that the unresolvable background from BBHs will be several orders of magnitude smaller than the corresponding background from BNS.  Therefore, we assume that the unresolvable  \ac{sgwb} arising from the BBH population is negligible and limit ourselves to the NSBH and BNS populations.

Since XG detectors will likely be able to detect all compact binaries that merge at low redshifts, such events will not contribute to the unresolved CBC SGWB. On the other hand, mergers at high redshifts will often not be individually resolvable, making them important contributors to the unresolved CBC SGWB. Therefore, we incorporate realistic star-formation-based redshift distribution models paired with a model of merger delay-time distributions. This is described further in Sec.~\ref{sec: redshift_distribution}. 

For all considered astrophysical populations, we assume that tidal effects and any eccentricity effects are negligible as they will be subdominant. The GW signal of such binaries can then be described by a set of 15 source parameters $\theta$, of which 8 are intrinsic and 7 are extrinsic. Intrinsic parameters include the component masses $m_1$ and $m_2$ and the three-dimensional spin vectors $\vec{\chi}_{1}$ and $\vec\chi_{2}$. Since spins are expected to have a subdominant effect on the SGWB, we also set them to zero~\cite{Zhou:2022nmt, Safarzadeh:2020qru}. Extrinsic parameters include the redshift \unboldmath $z$, the right ascension and declination ($\alpha, \delta$), the polarization angle $\psi$, the inclination angle $\iota$, the coalescence phase $\phi_c$, and the coalescence time $t_c$~\footnote{Any choice of $t_c$ does not affect the calculation of $\Omega_{\text{cbc, unres}}$, since it depends only on the GW power.}. We draw $\text{cos}\; \iota,\: \text{cos}\; \delta,\: \alpha,\: \psi,\: \text{and} \: \phi_c$ from uniform distributions.

The rate and mass models are described in the following subsections and the full distributions are summarized in Tab.~\ref{tab:1}.

\subsection{The redshift distributions}
\label{sec: redshift_distribution}

Compact binaries experience a delay time $t_d$ between their formation at $z_f$ and merger at $z$, where $z_f$ is the zero-age main-sequence (ZAMS) ``formation" redshift. We calculate the delay time as the difference between the cosmological lookback time~\cite{Hogg:1999ad} at $z_f$ and $z$,
\begin{equation}
\label{eqn: delay_time_def}
    t_d = t_L(z_f) - t_L(z),
\end{equation}

We assume a delay time distribution of $p(t_d) \propto t_d^{-1}$ as suggested by stellar evolution and population synthesis models~\cite{Piran:1992, Sana:2012px}. The GW and Galactic pulsar observations~\cite{Fishbach:2021mhp, Beniamini:2019iop} are also consistent with such a distribution, although Galactic populations likely harbor an excess of sources that will merge rapidly. An analysis using localized short gamma-ray bursts potentially finds steeper time-delay distributions, but it does not include selection effects~\cite{Zevin:2022dbo}. We set the maximum delay time $t_d^{max}$ to the Hubble time to limit ourselves to binaries that will merge in the age of the Universe. We set a fiducial minimum delay time $t_d^{min}$ of 20 Myr, which is approximately how long massive binaries take to evolve into two neutron stars~\cite{Regimbau:2012ir, Belczynski:2000wr, Belczynski:2006br}. It is also consistent with observations of binary pulsar merger times and of short gamma-ray bursts in both late- and early-type galaxies~\cite{Regimbau:2012ir, Berger:2006ik}. 

We convolve $t_d$ with a star formation rate (SFR) model $R_f(z_f)$ to calculate the source-frame CBC merger rate density per comoving volume~\cite{Fishbach:2021mhp}
\begin{equation}
\label{eqn: convolution}
    R_m(z) \propto \int_{t_d^{min}}^{t_d^{max}} R_f(\tilde{z}[t_L(z) + t_d]) p(t_d) dt_d,
\end{equation}
where $\tilde{z}$ is the formation redshift.
We normalize Eq.~\ref{eqn: convolution} as
\begin{equation}
\label{eqn: cbc_rate_normalization}
R_{\nu,cbc}(z) = \frac{\mathcal{R}_0}{R_m(z=0)} R_m(z), 
\end{equation}
such that $R_{\nu, cbc}(z=0) = \mathcal{R}_0$ is the inferred local source-frame CBC merger rate density per comoving volume obtained from population analysis (see Sec.~\ref{sec:masses_and_rates}) for CBCs of type $\nu$. Using Eq.~\ref{eqn: observed_rate_converted}, we calculate the observed differential CBC merger rate
\begin{equation}
\label{eqn: cbc_rate_basic}
\frac{dR_{o,cbc}(z)}{dz} = R_{\nu,cbc}(z) \frac{1}{1+z} \frac{dV_c}{dz} (z).
\end{equation}
We simulate CBCs up to redshift $z=10$ as we expect minimal star formation beyond this redshift and therefore no CBCs~\cite{Callister:2020arv}. Our fiducial SFR model is the Madau-Fragos SFR~\cite{Madau:2016jbv},
\begin{equation}
\label{eqn: mf_sfr}
R_f(z_f) \propto \frac{(1+z_f)^{2.6}}{1+(\frac{1+z_f}{3.2})^{6.2}}.
\end{equation}

\subsection{The mass distributions and rates}
\label{sec:masses_and_rates}

We now describe the mass distributions and local merger rate densities $\mathcal{R}_0$ we use to simulate the CBC population. We consider several population models, fit to LVK data up until the end of the third observing run via Bayesian inference. We also consider models of mass distribution fit to Galactic double neutron stars. Considering several models enables us to characterize systematic modeling uncertainties that could come from using strongly parameterized models. Each model has a statistical uncertainty on the population that we marginalize over. Additionally in each model, a single set of hyperparameters $\Lambda_{\text{GW}}$ characterizes a single CBC population. We assume that only the overall rate, not the mass distribution, evolves with redshift according to Eq.~\ref{eqn: convolution}.

For each model, we self-consistently include the rates inferred by it when available~\cite{Maya_Reed_Rates}. We refer to such models as ``rate-inclusive" models. For models where such information is not available -- which we refer to as ``rate-marginalized" models -- we randomly pair each hyperparameter sample with a $\mathcal{R}_0$ posterior from an LVK O3b population model~\cite{KAGRA:2021duu} that estimates rates. For clarity, we rewrite Eq.~\ref{eqn: cbc_rate_basic} for cases where $\mathcal{R}_0$ depends on $\Lambda_{\text{GW}}$ (i.e., $\mathcal{R}_0 (\Lambda_{\text{GW}})$) as
\begin{equation}
\label{eqn: cbc_rate_revised}
\frac{dR_{o,cbc}(z, \Lambda_{\text{GW}})}{dz} = R_{v,cbc}(z, \Lambda_{\text{GW}}) \frac{1}{1+z} \frac{dV_c}{dz} (z).
\end{equation}
The various mass-rate model configurations are described below and summarized in Tab.~\ref{tab:1}.

\begin{table*}[t]
    \centering
	\begin{tabular}{|P{0.16\textwidth}|P{0.16\textwidth}|P{0.16\textwidth}|P{0.16\textwidth}|P{0.16\textwidth}|P{0.16\textwidth}|} 
    \hline 
    \multicolumn{1}{|c|}{}  & \multicolumn{5}{c|}{\textbf{Model}} \\ 
    \hline
    \textbf{Parameter}                & \textsc{pdb-pl}      & \textsc{nsbh-pl}       & \textsc{nsbh-g}  & \textsc{bns-g}      & \textsc{bns-pl} \\
    \hline
    \boldmath $\mathcal{R}_0 (\Lambda_{\text{GW}})$ & Rate-inclusive & Rate-inclusive  & Rate-inclusive  &  Drawn from \textsc{pdb-ind} analysis~\cite{Farah:2021qom, Fishbach:2020ryj, KAGRA:2021duu, LIGOScientific:2021djp} &Drawn from \textsc{pdb-ind} analysis~\cite{Farah:2021qom, Fishbach:2020ryj, KAGRA:2021duu, LIGOScientific:2021djp}\\ 
    \hline
    \boldmath ${m}_{1}$        & PDB with power-law pairing~\cite{Farah:2021qom, KAGRA:2021duu, LIGOScientific:2021djp} & Power Law~\cite{Biscoveanu:2022iue} & Power Law~\cite{Biscoveanu:2022iue} & Double Gaussian~\cite{Farrow:2019xnc} & Power Law~\cite{KAGRA:2021duu, Landry:2021hvl, LIGOScientific:2021djp}\\ 
    \hline
    \boldmath ${m}_{2}$        & PDB with power-law pairing~\cite{Farah:2021qom, KAGRA:2021duu, LIGOScientific:2021djp} & Power Law Pairing~\cite{Biscoveanu:2022iue} & Gaussian Pairing~\cite{Biscoveanu:2022iue} & Uniform~\cite{Farrow:2019xnc} & Random Pairing~\cite{KAGRA:2021duu, Landry:2021hvl, LIGOScientific:2021djp}\\ 
    \hline
    \multicolumn{1}{|c|}{\textbf{\boldmath $z$}}  & \multicolumn{5}{c |}{Madau-Fragos SFR~\cite{Madau:2016jbv}} \\ 

    \multicolumn{1}{|c|}{\textbf{\boldmath $t^{min}_{d}$}}  &\multicolumn{5}{c|}{20 Myr}\\ 
  
    \multicolumn{1}{|c|}{\textbf{\boldmath $\vec{\chi}_{1}$, $\vec{\chi}_{2}$}}  &\multicolumn{5}{c |}{0} \\ 
    
    \multicolumn{1}{|c|}{\textbf{\boldmath cos $\iota$, cos $\delta$}}  &\multicolumn{5}{c |}{Uniform in $[-1, 1]$}\\ 
    
    \multicolumn{1}{|c|}{\textbf{\boldmath $\alpha$, $\psi$, $\phi_{c}$}}  &\multicolumn{5}{c|}{Uniform in $[0, \boldmath 2\pi]$}\\ 
     
    \hline 
    \end{tabular}
    
\caption{Different populations of the various source parameters used in calculating the  \ac{sgwb} from CBCs. See Section \ref{sec:pop_models} for further details.}
\label{tab:1}
\end{table*}

\subsubsection{Neutron-star black-hole models}
\label{sec:nsbh_mass_models}

We model the NSBH mass distribution using the Bayesian population analyses from Ref.~\cite{Biscoveanu:2022iue}. We assume that the distribution of the primary mass ${m}_{1}$ follows the truncated power-law~\cite{Fishbach:2017zga}
\\ \\
$\pi(m_1|\gamma, m_{1,\text{min}}, m_{1,\text{max}}) \propto$
\begin{equation}
\label{eqn:nsbh_m1}
    \begin{cases}
        m_1^{-\gamma}, \; \text{if} \; m_{1, \text{min}} \leq m_1 \leq m_{1, \text{max}}\\
        0, \; \text{otherwise}
    \end{cases},
\end{equation}
with a power-law index $\gamma$, minimum $m_1$ cutoff $m_{1,\text{min}}$, and maximum $m_1$ cutoff $m_{1,\text{max}}$.

We consider two pairing functions to get the distribution of the secondary mass. The first is a power-law pairing function~\cite{Fishbach:2019bbm}, which we refer to as the \textsc{nsbh-pl} model
\\ \\
$\pi(q|\beta, m_1, m_{2,\text{max}}) \propto$
\begin{equation}\label{eqn:nsbh_power}
    \begin{cases}
        q^{\beta}, \; \text{if} \; q_{\text{min}}(m_1) \leq q \leq  q_{\text{max}}(m_1, m_{2,\text{max}}))\\
        0, \; \text{otherwise}
    \end{cases},
\end{equation}
where $q = m_2/m_1$ is the mass ratio and $\beta$ is a power-law index. The second is a truncated-Gaussian pairing function~\cite{Biscoveanu:2022iue}, which we refer to as the \textsc{nsbh-g} model
\\ \\
$\pi(q|\mu, \sigma, m_1, m_{2,\text{max}}) \propto$
\begin{equation}
\label{eq:nsbh_gaussian}
    \begin{cases}
        \mathcal{N}(q|\mu, \sigma), \; \text{if} \; q_{\text{min}}(m_1) \leq q \leq  q_{\text{max}}(m_1, m_{2,\text{max}}))\\
        0, \; \text{otherwise}
    \end{cases},
\end{equation}
where $\mathcal{N}(q|\mu, \sigma)$ is a Gaussian with mean $\mu$ and standard deviation $\sigma$. For both models, we set the minimum NS mass $m_{2,\text{min}} = 1 \ M_\odot$ so that the minimum mass ratio cutoff $q_{\text{min}} = 1/m_1$. The maximum mass ratio cutoff is set as $q_{\text{max}} = \text{min}(m_{2,\text{max}}/m_1, 1)$, where the maximum NS mass $m_{2,\text{max}}$ is a free parameter drawn uniformly between $1.97 \ M_\odot$ and $2.7 \ M_\odot$. While the ranges of $m_1$ and $m_2$ differ based on the particular hyperparameter values, the broadest possible range is $m_1 \in [2, 20] \ M_\odot$ and $m_2 \in [1, 2.7] \ M_\odot$. We refer the reader to Ref.~\cite{Biscoveanu:2022iue} for a more detailed overview of the models.

Both the \textsc{nsbh-pl} and \textsc{nsbh-g} models are rate-inclusive models that self-consistently calculate $\mathcal{R}_0$. Hence, for each hyperparameter sample in both analyses, we set $\mathcal{R}_0$ to the associated posterior $\mathcal{R}_0(\Lambda_{\text{GW}})$.

\subsubsection{Power Law + Dip + Break model}
\label{sec:joint_mass_models}

We next consider the Power Law + Dip + Break (PDB) model used in the LVK GWTC-3 analysis~\cite{Farah:2021qom, Fishbach:2020ryj, KAGRA:2021duu, LIGOScientific:2021djp}. This model fits the entire mass spectrum of CBCs eschewing the difference between black holes and neutron stars. The PDB model fits both the primary and secondary masses using a common distribution $\pi_{\rm pdb}(m|\vec{\Lambda})$ that is based on a broken-power law and includes a ``notch filter" to model a potential mass gap between $3 \text{ to } 5 M_{\odot}$. We refer the reader to Ref.~\cite{Farah:2021qom} for a detailed overview of the model.

We then use a power-law pairing function to pair the primary and secondary masses~\cite{Fishbach:2019bbm, KAGRA:2021duu}:
\begin{equation}\label{eqn:pdb_model_pairing}
    g(m_1, m_2, \beta) =
    \begin{cases}
        q^{\beta}, \; \text{if} \; m_2 \leq m_1\\
        0, \; \text{otherwise}
    \end{cases}.
\end{equation}
This gives the following joint mass distribution, which we refer to as the \textsc{pdb-pl} model:
\begin{equation}
\label{eqn:pdb_model_joint}
        \pi_{\textsc{pdb-pl}}(m_1, m_2| \vec{\Lambda}, \beta) \propto 
        \pi_{\rm pdb}(m_1|\vec{\Lambda})\pi_{\rm pdb}(m_2|\vec{\Lambda}) g(m_1, m_2, \beta) \;.
\end{equation}

While \textsc{pdb-pl} is a rate-inclusive model, we need to account for the fact that the model natively incorporates BBH mergers as well. In order to calculate rates that only include the NSBH and BNS parts of the spectrum, we correct the rates by the fraction of NSBH and BNS mergers that this model predicts (at each point in the hyperparameter space). Since we need systems to have at least one neutron star, we limit $m_2$ to span $[1, 3] \: M_{\odot}$ while allowing $m_1$ to span $[1, 100] \: M_{\odot}$ as defined in the original model~\cite{KAGRA:2021duu}.

\subsubsection{Binary neutron star models}
\label{sec:bns_mass_models}

We model the BNS mass distribution using two different rate-marginalized models. The first model is motivated through observations of Galactic double neutron stars, which provide the largest and most well-constrained sample of such systems. This model, which we refer to as \textsc{bns-g}, is based on the most favored model from the population analysis of the 17 observed Galactic BNSs by Ref.~\cite{Farrow:2019xnc}. In particular, the primary neutron-star mass follows a double-Gaussian mixture model

\begin{equation}
\label{eqn:bns_galactic_m1}
\begin{split}
    \pi(m_1|\gamma, \mu_1, \sigma_1, \mu_2, \sigma_2) = & \frac{\gamma}{\sigma_1 \sqrt{2\pi}} {\rm e}^{-\frac{(m_1 - \mu_1)^2}{2\sigma_1}}  \\
     & + \frac{1 - \gamma}{\sigma_2 \sqrt{2\pi}} {\rm e}^{-\frac{(m_2 - \mu_2)^2}{2\sigma_2}},
\end{split}
\end{equation}
where $\gamma$ ($1 - \gamma$), $\mu_1$ ($\mu_2$), and $\sigma_1$ ($\sigma_2$) give the weight, mean, and standard deviation of the first (second) peak, respectively. The secondary neutron-star mass is drawn uniformly within the range
\begin{equation}
\label{eqn:bns_galactic_m2}
\pi(m_2|m_{2, \text{min}}, m_{2, \text{max}}) = U[m_{2, \text{min}}, m_{2, \text{max}}],
\end{equation}
with minimum and maximum cutoff $m_{2,min}$ and $m_{2,max}$, respectively. As usual, we assume that $m_2 \leq m_1$. We note that both the double-Gaussian and uniform distributions are truncated at 0.8 $M_\odot$ and 2 $M_\odot$ to reflect the prior ranges of the analysis from Ref.~\cite{Farrow:2019xnc}.

While the analysis in Ref.~\cite{Farrow:2019xnc} relies on observations of only Galactic BNSs, we assume in our usage of the \textsc{bns-g} model that it is extendable to all redshifts and metallicities. Note, however, that the BNS mass distribution inferred with Galactic observations is potentially inconsistent with GW observations. This is especially highlighted by the BNS merger GW190425~\cite{LIGOScientific:2020aai}, which has a total mass heavier than that of typical Galactic double neutron stars and could potentially have formed through different formation channels~\cite{Romero-Shaw:2020aaj}. 

Hence, we include a second model, which we call \textsc{bns-pl}, based on the BNS \textsc{power} mass distribution inferred from all the neutron star systems detected through GWs~\cite{LIGOScientific:2021djp, KAGRA:2021duu, Landry:2021hvl}. In the \textsc{bns-pl} model, both the primary and secondary masses are paired randomly after each being drawn from a power-law distribution
\begin{equation}
\label{eqn:bns_ns_power}
\pi(m|\gamma, m_{\text{min}}, m_{\text{max}}) \propto m^\gamma,
\end{equation}
with a power-law index $\gamma$ and minimum (maximum) mass cutoff $m_{\text{min}}$ ($m_{\text{max}}$). While the neutron star \textsc{power} model is used in Ref.~\cite{ KAGRA:2021duu} to infer the mass distribution from both BNSs and NSBHs, we only consider the BNS case here.

Since our BNS models are all rate-marginalized, we model their rate distributions by drawing $\mathcal{R}_0$ posteriors from the rate-inclusive PDB random-pairing model (which we refer to as \textsc{pdb-ind})~\cite{Farah:2021qom, KAGRA:2021duu} fit to LVK GWTC-3 data. Similar to Sec.~\ref{sec:joint_mass_models}, we correct the rates inferred by \textsc{pdb-ind} to include only the fraction of systems that are BNSs. This means restricting the masses to $m_1, m_2 \in [1, 3] \: M_{\odot}$ for the \textsc{bns-pl} model, but to $m_1, m_2 \in [1, 2] \: M_{\odot}$ for the \textsc{bns-g} model to be consistent with the mass ranges from Refs.~\cite{Farrow:2019xnc, KAGRA:2021duu}.

\begin{figure*}[ht]
    \centering
    \includegraphics[width=0.80\textwidth]{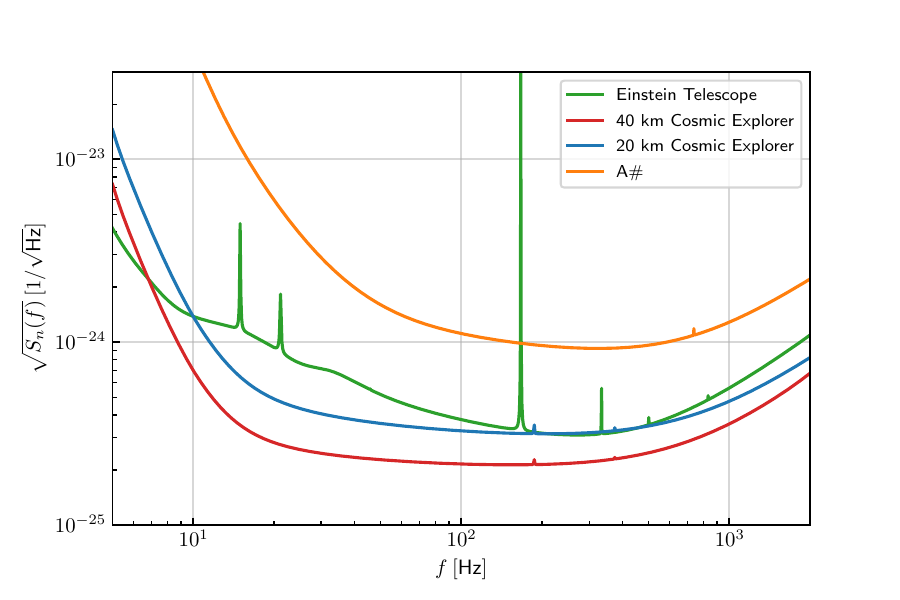}
    \caption{
    Amplitude spectral densities $\sqrt{S_n}$ for the detectors in our networks.
    }
    \label{fig:psd}
\end{figure*}

\section{Detector Networks}
\label{sec:detector_networks}

We consider four different possible networks of XG detectors, using the latest detector designs described in Refs.~\cite{Evans:2023euw, Gupta:2023lga}. Using the GW simulation package \textsc{Gwbench}~\cite{Borhanian:2020ypi}, we consider the \texttt{CE-40} and \texttt{CE-20} options for the proposed 40-km arm and 20-km arm \ac{ce} sensitivities respectively~\cite{Evans:2023euw, Gupta:2023lga}, the \texttt{A\#} option for the proposed 4-km arm $\text{A}^{\#}$ sensitivity~\cite{A_Sharp_sensitivity}, and the \texttt{ET-10-XYL} option for the proposed triangular ``xylophone" 10-km arm \ac{et} sensitivity~\cite{ET_10km_sensitivity}.  Figure~\ref{fig:psd} shows the amplitude spectral density $\sqrt{S_n (f)}$ of each detector considered.

Since an \ac{sgwb} is detected by cross-correlating data observed separately by at least two detectors, we consider only multiple-detector networks~\cite{Allen:1997ad}. We consider the \texttt{CE-A} (coast of Washington, USA), \texttt{CE-B} (coast of Texas, USA), \texttt{ETS} (slightly South of Virgo's current location at Cascina, Italy), and \texttt{LLO} (current location of the LIGO-Livingston Observatory at Livingston, Louisiana, USA) facility locations specified in Tab.~II of Ref.~\cite{Gupta:2023lga}, which we point to for the details.
\pagebreak

We consider four different networks of detectors:
\begin{itemize}

    \item A \textit{fiducial} three-detector network including one \texttt{CE-40} at \texttt{CE-A}, one \texttt{CE-20} at \texttt{CE-B}, and one \texttt{ET-10-XYL} at \texttt{ETS}.

    \item An \textit{alternate} three-detector network including one \texttt{CE-40} at \texttt{CE-A}, one \texttt{ET-10-XYL} at \texttt{ETS}, and one \texttt{A\#}-upgraded detector at \texttt{LLO}.

    \item An \textit{alternate} two-detector network including one \texttt{CE-40} at \texttt{CE-A}, and one \texttt{CE-20} at \texttt{CE-B}.
    
    \item An \textit{alternate} two-detector network including one \texttt{CE-40} at \texttt{CE-A}, and one \texttt{ET-10-XYL} at \texttt{ETS}.
\end{itemize}

We choose the same three-detector networks as those proposed in Refs.~\cite{Evans:2023euw, Gupta:2023lga} to make it easy to compare with existing and future literature. While the future development of \ac{et} has been officially confirmed, we specifically explore a two-detector network configuration that does not include \ac{et} in order evaluate the usefulness of and aid in the detector design of \ac{ce} facilities in detecting astrophysically and cosmologically-arising \acp{sgwb} even in the absence of ET.

Throughout the paper, we set a minimum frequency of 5 Hz, corresponding to the proposed lower limit of CE and $\text{A}^{\#}$ design sensitivities~\cite{Evans:2023euw, Gupta:2023lga, A_Sharp_sensitivity} as shown in Fig.~\ref{fig:psd}. In addition, we set a maximum frequency of 2000 Hz. For each network, we use the same $3\sigma$ power-law integrated (PI) curves~\cite{Thrane:2013oya, Zhou:2022nmt} as Ref.~\cite{Gupta:2023lga}~\footnote{Obtained through private communication with the authors.} in order to measure the ability of the network to detect \ac{sgwb} signals. We refer the reader to Ref.~\cite{Gupta:2023lga} for the overlap reduction functions~\cite{Thrane:2013oya, Zhou:2022nmt} of the various detector pairs comprising the networks described in this section.


\section{Simulating the SGWB : A Monte-Carlo approach}
\label{sec:methods_2}

In the previous sections, we have described the population models and the detector networks that we use. In this section, we describe the simulation and the calculation of the unresolved \ac{sgwb} using Eq.~\ref{eqn: omega_gw_final} and Eq.~\ref{eqn: spectral_energy_flux_conversion} for which we adopt a Monte-Carlo approach~\cite{Regimbau:2022mdu, Regimbau:2016ike, Regimbau:2012ir, Regimbau:2014uia, Meacher:2015iua}. We use the \textsc{Gwbench} simulation platform~\cite{Borhanian:2020ypi} to generate GW waveforms for non-spinning quasi-circular binaries neglecting tidal effects. \textsc{Gwbench} naturally accounts for the Earth's rotation and its impact on the antenna patterns which is important since BNSs can last for several hours in the observing band of XG detectors~\cite{Borhanian:2020ypi}. We use IMRPhenomD~\cite{Husa:2015iqa, Khan:2015jqa} as our waveform approximant~\footnote{Note that the choice of waveform approximant is expected to be subdominant in calculating the SGWB~\cite{Zhou:2022nmt}.}.

For each population configuration described in Sec.~\ref{sec:pop_models} and summarized in Tab.~\ref{tab:1}, we are interested in calculating $\Omega_{\text{cbc, unres}}$. To account for astrophysical uncertainties, we draw  2000 hyperparameter samples from the inferred population distributions and estimate $\Omega_{\text{cbc, unres}}$ for each sample.

To do this in a computationally tractable way, we first simulate $10^5$ waveforms each for NSBHs and BNSs drawn from broad fiducial population distributions~\footnote{In this fiducial population, for NSBHs we uniformly draw $m_1 \sim [2, 50]$ and $m_2 \sim [1, 3]$ and for BNSs we uniformly draw both $m_1, m_2 \sim [0.5, 3]$.}. We then apply rejection sampling to draw a population corresponding to any particular hyperparameter draw. We estimate the mean number of sources needed for a CBC model $\nu$ with a set of hyperparameters $\Lambda_{\text{GW}}$ and observation time $T$ as

\begin{equation}
\label{eqn: expected_sources_comoving}
        \langle N_{\nu}(\Lambda_{\text{GW}}) \rangle =  \int^{z_{\text{max}}}_{0} R_{\nu}(z, \Lambda_{\text{GW}}) \frac{dV_c}{dz} \frac{dz}{1+z} \; T,
\end{equation}
and draw the actual number of sources through a Poisson draw

\begin{equation}
\label{eqn: observed_sources}
N_{\nu}(\Lambda_{\text{GW}}) \sim  \texttt{Poiss} ( \lambda = \langle N_{\nu}(\Lambda_{\text{GW}}) \rangle).
\end{equation}

In order to estimate $\Omega_{\text{cbc, unres}}$, we first need to extract only the unresolvable signals from this population. To do this, we first compute the matched-filter signal-to-noise ratio (SNR) $\rho_j^{\rm mf}$ of the signal, defined for detector $j$ with noise PSD $S_{n, j} (f)$ as~\cite{Moore:2014lga}: 

\begin{equation}
\label{eqn: matched_filter_snr}
\rho_j^{\rm mf} = \left[ 4 \int_{0}^{\infty} \frac{|\tilde{h}_j (f)|^2}{S_{n,j}(f)} df \right]^{1/2}. 
\end{equation}
This is, however, the optimal matched-filter SNR; in order to account for the measurement uncertainty due to detector noise, we correct the SNR as~\cite{Fishbach:2019ckx, Callister:2020arv, Fishbach:2018edt}

\begin{equation}
\label{eqn: observed_snr}
\rho_j^{\rm obs} = \rho_j^{\rm mf} + \mathcal{N}(0, 1).
\end{equation}
We then define the optimal network SNR $\rho_{\text{net}}^{\rm obs}$ for a network of $D$ detectors by summing the individual observed SNRs in quadrature as
\begin{equation}
\label{eqn: network_snr}
\rho_{\text{net}}^{\rm obs} = \sqrt{\sum^{D}_{j=1} (\rho^{\rm obs}_j)^2}.
\end{equation}
A signal is then labeled as resolved if $\rho_{\text{net}}^{\rm obs} $ is greater than some frequency-independent threshold $\rho_{\text{thresh}}$.

Once we have a population $N_{\nu}^{\rm unres}(\Lambda_{\text{GW}})$ of unresolvable sources with respect to $\rho_{\text{thresh}}$, we calculate the GW energy flux (Eq.~\ref{eqn: omega_gw_flux_def}) using a Monte-Carlo sum over our simulated population~\cite{Zhou:2022nmt, Regimbau:2016ike, Meacher:2015iua}:
\begin{equation}
\label{eqn: total_flux_numerical}
F_{\nu}(f; \Lambda_{\text{GW}}) \approx \frac{\pi c^3 f^2}{2G T} \sum_{i=1}^{N_{\nu}^{\text{unres}}(\Lambda_{\text{GW}})} \biggr[|\tilde{h}_{+}^i(f,\theta^i)|^2 + |\tilde{h}_{\times}^i(f,\theta^i)|^2 \biggr].
\end{equation}

\begin{figure*}[hbtp]
    \centering
    \begin{subfigure}[b]{0.9\textwidth}
            \includegraphics[width=0.9\textwidth]{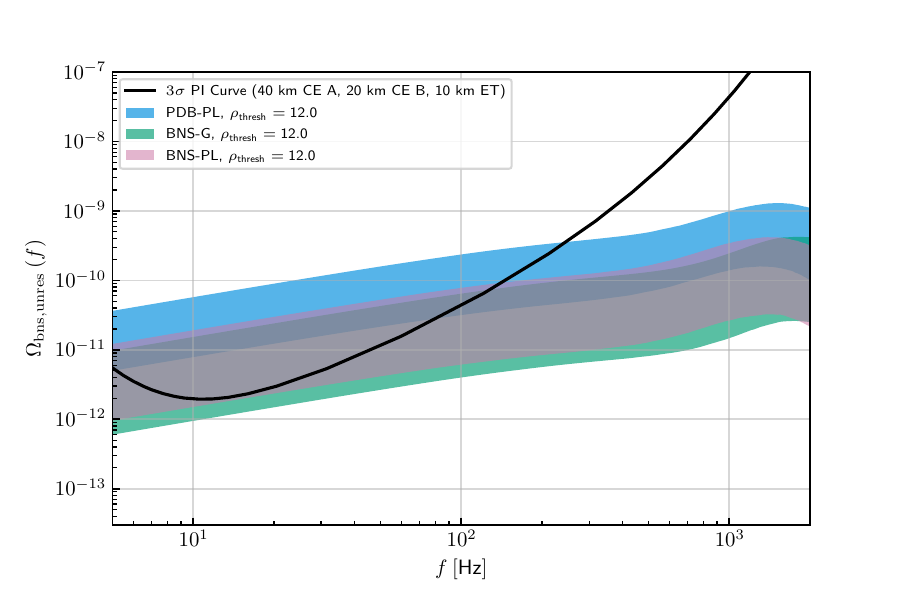}
            \caption{$\Omega_{\text{cbc, unres}}$ estimates for the various BNS models in Tab.~\ref{tab:1}. The bands show the central $90\%$ credible levels.}
            \label{fig:1CE40_1CE20_bns_envelope}
    \end{subfigure}

        \begin{subfigure}[b]{0.9\textwidth}
            \includegraphics[width=0.9\textwidth]{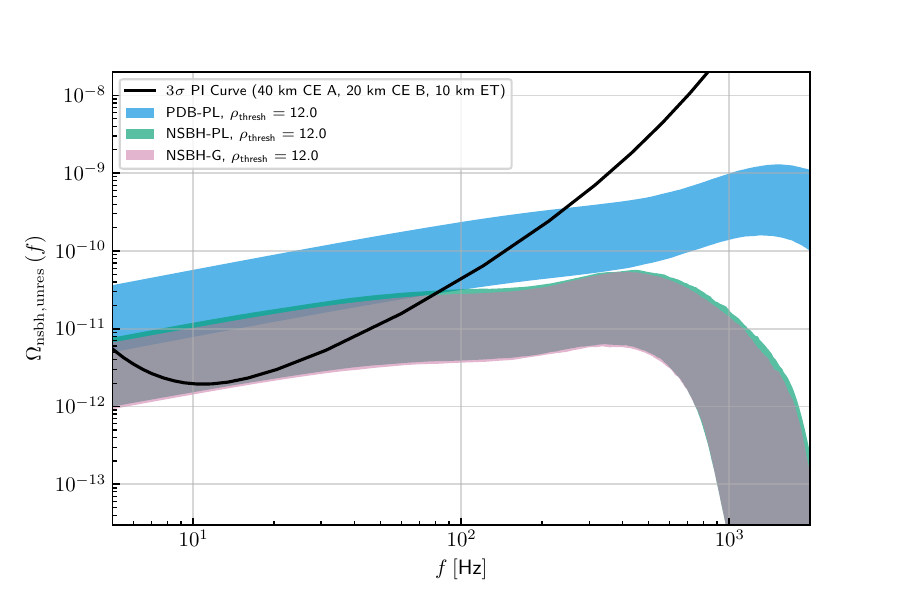}
            \caption{$\Omega_{\text{cbc, unres}}$ estimates for the various NSBH models in Tab.~\ref{tab:1}. The bands show the central $90\%$ credible levels.}
        \label{fig:1CE40_1CE20_nsbh_envelope}
    \end{subfigure}
        \caption{This plot shows the $\Omega_{\text{cbc, unres}}$ estimate for the various BNS and NSBH models for the fiducial 3-detector network described in Sec.~\ref{sec:detector_networks}. The purple band in both plots is the \textsc{pdb-pl} model that simultaneously incorporates both the NSBH and BNS systems. Also overlaid is the $3\sigma$ PI curve that shows that the unresolved CBC backgrounds will likely be very loud.}
    \label{fig:bns_and_nsbh}
\end{figure*}

\begin{figure*}[ht]
    \centering
    \includegraphics[width=0.85\textwidth]{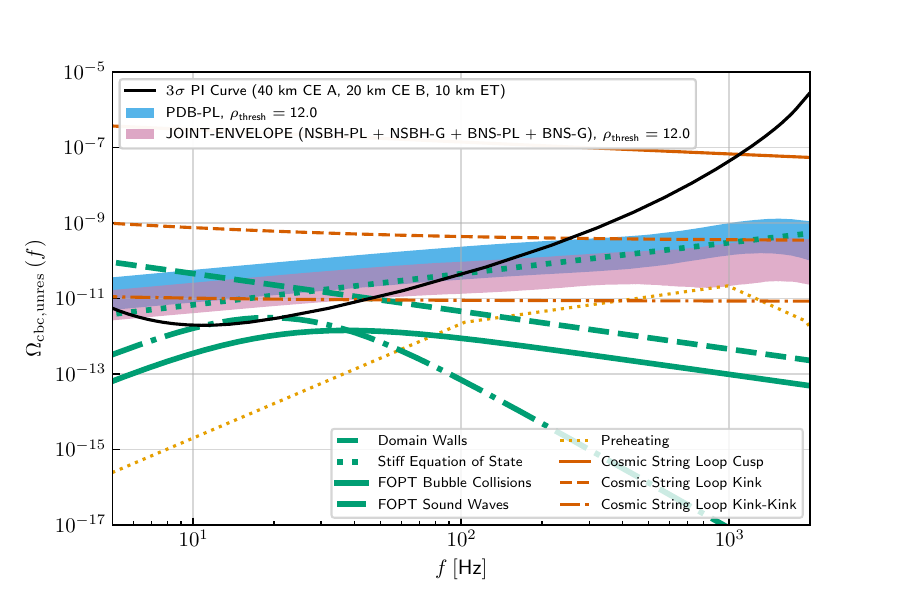}
    \caption{A comparison of various cosmological \ac{sgwb} models along with $\Omega_{\text{cbc, unres}}$ estimates for the fiducial network described in Sec.~\ref{sec:detector_networks}. The blue band shows the \textsc{pdb-pl} model while the pink band shows the \textsc{joint-envelope} constructed from the various BNS-only and NSBH-only models in Tab.~\ref{tab:1}. Both bands represent $90\%$ credible intervals. The solid black curve is the $3 \sigma$ PI curve.}
    \label{fig:fiducial_joint_unresolved}
\end{figure*}

\section{Cosmological SGWB Models}
\label{sec:cosmo_backgrounds}

To gauge the impact of the $\Omega_{\text{cbc, unres}}$, we consider several models of early Universe cosmological \acp{sgwb} that could be accessible with XG detectors in principle. These models include \acp{sgwb} from domain walls~\cite{Saikawa:2017hiv}, first-order phase transition (FOPT) sound waves~\cite{Romero:2021kby, Espinosa:2010hh}, FOPT bubble collisions~\cite{Romero:2021kby}, stiff equation of state~\cite{Figueroa:2019paj}, and preheating~\cite{Easther:2006vd, Gupta:2023lga}. We also use the model spectrum for Nambu-Goto oscillating cosmic string loops (the \textsc{model C1} from Ref.\cite{LIGOScientific:2021nrg}).

We note that XG detectors will not be sensitive to standard slow-roll inflation, for which $\Omega_{\text{GW}} \sim 10^{-15} - 10^{-17}$ depending on the model~\cite{Caprini:2018mtu, Zhou:2022otw}. In general, these various \acp{sgwb} are highly sensitive to the choice of model parameters, but in our plots, we chose illustrative curves for $\Omega_{\rm cosmo}$.

\section{Results} 
\label{sec:results}

As described in Sec.~\ref{sec:methods_2}, for each population model in Tab.~\ref{tab:1}, we have 2000 different draws for the unresolved CBC \ac{sgwb}. These draws incorporate both the uncertainty in the rate of mergers and the astrophysical uncertainty from the population model. In Fig.~\ref{fig:bns_and_nsbh}, we show the $90\%$ credible bands for the unresolved \ac{sgwb} for the various BNS-only and NSBH-only population models, for the fiducial 3-detector network from Sec.~\ref{sec:detector_networks} using a threshold SNR of 12. We overlay them on top of the $3 \sigma$ PI curve for this network. We assume an observation time of $T = 1$ year throughout the paper. The purple band in both plots is from the $\textsc{pdb-pl}$ model and is therefore higher than either of the BNS-only and NSBH-only models. 

The unresolved \ac{sgwb} from CBCs is above the $3 \sigma$ PI curve, implying that it will very likely be detectable by XG detectors with one year of observing. In general, the width of each band in Fig.~\ref{fig:bns_and_nsbh} -- which represents the total uncertainty in our estimate of the unresolved \ac{sgwb} for that model --  spans about an order of magnitude and the bands fall out of detectability at $\sim 80$ Hz even with XG detectors. The combined \textsc{pdb-pl} background, however, might be significantly detectable up to $\sim 200$ Hz. Nevertheless, it would seem that we will only observe the \ac{sgwb} due to the inspiral phase of the mergers. Similarly, while the simulations show a clear turnover for the NSBH \ac{sgwb} -- the morphology of which probably depends on the waveform used -- it clearly happens at high frequencies and XG detectors will probably not be able to probe this. 

\begin{figure*}[ht]
    \centering
    \includegraphics[width=0.85\textwidth]{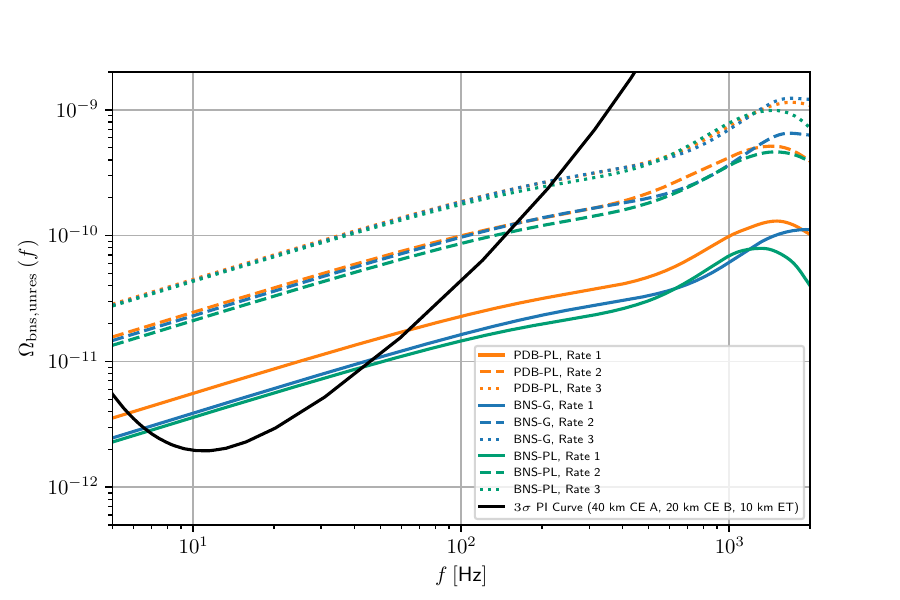}
    \caption{The \ac{sgwb} for different BNS models evaluated at three different rates using a threshold SNR of 12. We use the following NS-system rates (corrected BNS rates for \textsc{bns-pl} and \textsc{bns-g}): Rate 1 = 45.5 (32.2) $\text{Gpc}^{-3} \: \text{Yr}^{-1}$, Rate 2 = 209.3 (186.6) $\text{Gpc}^{-3} \: \text{Yr}^{-1}$, Rate 3 = 377.0 (362.6) $\text{Gpc}^{-3} \: \text{Yr}^{-1}$. While there is significant difference between the curves for various rates, the differences between the models is minimal.}
    \label{fig:1CE40_1CE20_mass_rate_comparison}
\end{figure*}

In addition to the \textsc{pdb-pl} model which naturally incorporates both BNS and NBSH mergers, we create a maximum-uncertainty envelope, which we refer to as the \textsc{joint-envelope} as a means of accounting for model systematics. This is generated by combining the BNS-only and NSBH-only models (i.e., \textsc{nsbh-pl}, \textsc{nsbh-g}, \textsc{bns-pl}, \textsc{bns-g}) in Tab.~\ref{tab:1} in all possible combinations and choosing the widest band at each frequency bin.

Figure~\ref{fig:fiducial_joint_unresolved} shows the $90 \%$ credible band of the \textsc{joint-envelope} along with the \textsc{pdb-pl} estimate of $\Omega_{\text{cbc, unres}}$ for our fiducial 3-detector network. We contrast these bands with several cosmological \ac{sgwb} models, demonstrating that the unresolved background will likely be a major source of noise for accessing the said cosmological signals. Since the effective CBC background is the sum of unresolved and resolved components (see Eq.~\ref{Eq:cbc_forground}), it also depends on the efficacy of subtraction of resolved signals. Since the subtraction residue could contribute significantly to the effective CBC background (see for e.g.~\cite{Zhong:2022ylh, Sharma:2020btq}), the unresolved background we show in Fig.~\ref{fig:fiducial_joint_unresolved} therefore represents the floor of the noise from CBC sources. Either way, a joint simultaneous analysis of the CBC background and the cosmological background will be necessary for the detection of the latter. We note again that the cosmological \ac{sgwb} curves themselves can change depending on the choice of parameters. In Appendix.~\ref{sec:appendix_A}, we present similar results for the three alternate detector networks described in Sec.~\ref{sec:detector_networks}.

Further exploring the uncertainty in the $\Omega_{\text{cbc, unres}}$, we see remarkable consistency between the two NSBH-only models, and similarly between the two BNS-only models. This suggests that the uncertainty in the rate is dominant, as opposed to the astrophysical uncertainty on the mass distribution. This is further confirmed by Fig.~\ref{fig:1CE40_1CE20_mass_rate_comparison}, where we plot the \ac{sgwb} from different BNS models for three different rates. The consistency in \ac{sgwb} between models at any given frequency shows that the astrophysical uncertainty is a much smaller contributor to the \ac{sgwb} uncertainty than the uncertainty in CBC rate. We find similar results for NSBHs as well. Since the number of BNSs and NSBHs observed thus far is fairly small~\cite{LIGOScientific:2018mvr, LIGOScientific:2020ibl, LIGOScientific:2021djp}, this means that any future detections in O4 and O5 can substantially lower the uncertainty of these bands. This is also consistent with studies such as Ref.~\cite{Jenkins:2018kxc}, which show that the monopole of the CBC \ac{sgwb} is particularly sensitive to the local merger rate. 

\begin{figure*}[ht]
    \centering
    \includegraphics[width=0.85\textwidth]{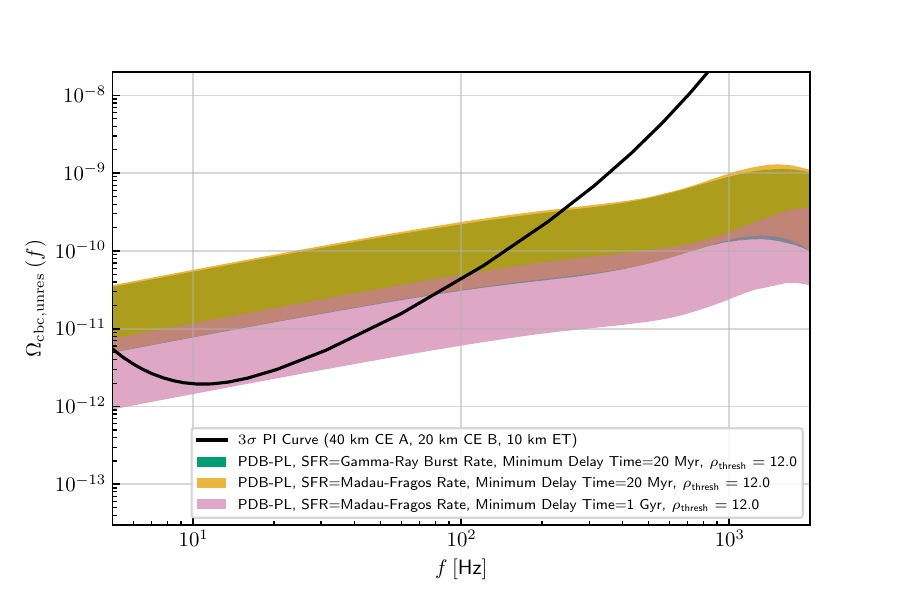}
    \caption{90\% credible intervals of $\Omega_{\text{cbc, unres}}$ for the \textsc{pdb-pl} model with three different SFR assumptions. While the green and the saffron bands are for a gamma-ray burst based SFR and the Madau-Fragos SFR respectively, the pink band is for a conservative 1 Gyr minimum time delay.}
    \label{fig:1CE40_1CE20_sfr_td_mdr}
\end{figure*}

To explore how SFR models impact our estimates of $\Omega_{\text{cbc, unres}}$, we also used an SFR extracted from long gamma-ray burst rates following~\cite{Zhou:2022nmt, Vangioni:2014axa}:

\begin{equation}
\label{eqn: grb_sfr}
R_f(z_f) \propto \nu \frac{ae^{b(z_f - z_p)}}{a - b + be^{a(z_f - z_p)}},
\end{equation}
where $\nu=0.146 \: M_\odot \: {\text{yr}}^{-1} \: {\text{Mpc}}^3$, $z_p = 1.72$, $a = 2.80$, and $b=2.46$.
As the \textsc{pdb-pl} bands in Fig.~\ref{fig:1CE40_1CE20_sfr_td_mdr} show, we find that the impact of the SFR model is minimal as long as the minimum delay times are small. A minimum delay time of $1$ Gyr gives a significantly smaller, albeit still loud, \ac{sgwb}. Note, however, that studies of Galactic systems have actually found an excess of systems with small delay times~\cite{Beniamini:2019iop}, meaning that the value of $1$ Gyr is very conservative. We therefore conclude that the key observation of this paper, that the unresolved background will significantly impact searches for cosmological backgrounds, is likely robust to uncertainties in SFR. We have also tested our assumption of $z_{\rm max} = 10$ and find that the GW power from higher redshifts is very small. 

The LVK population inference at the end of O3 tested multiple versions of the \textsc{power} model~\cite{KAGRA:2021duu} including only confident events, or confident and marginal events, or confident events and GW190814~\footnote{which is a population outlier~\cite{Essick:2021vlx}}. While we use the last version in all our plots as the \textsc{bns-pl} model, we find that the impact of this choice is negligible. In particular, our estimates of $\Omega_{\text{bns, unres}}$ for our implemented version differs from those for the other two versions at most by $3\%$ at 25 Hz, demonstrating that our estimates are robust with respect to this uncertainty.

\begin{figure*}[ht]
    \centering
    \includegraphics[width=0.85\textwidth]{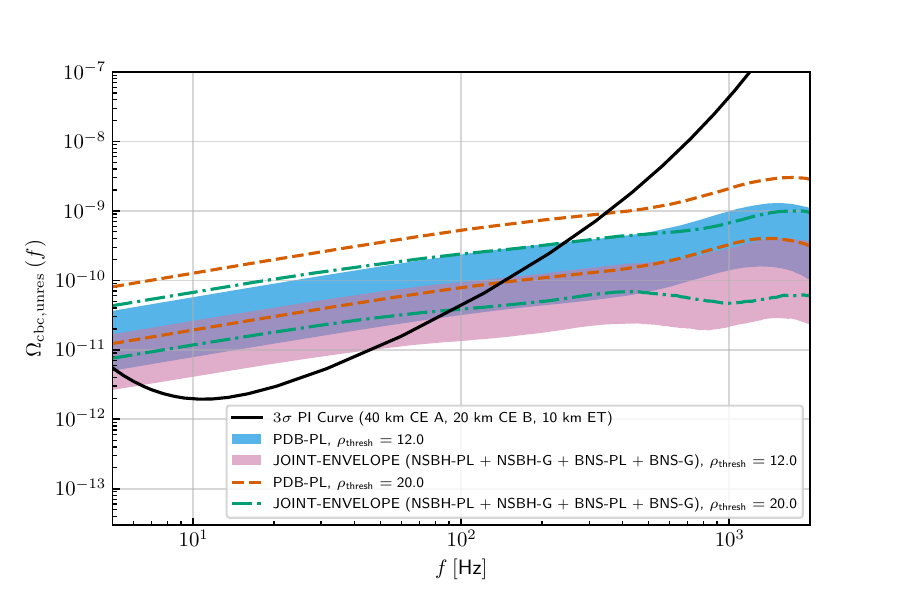}
    \caption{90\% credible intervals of $\Omega_{\text{cbc, unres}}$ for the \textsc{pdb-pl} and the \textsc{joint-envelope} models with two different $\rho_{\rm thresh}$ values. The filled bands show the credible intervals with $\rho_{\rm thresh} = 12$ and the dashed-dotted lines bound the intervals for $\rho_{\rm thresh} = 20$.}
    \label{fig:1CE40_1CE20_snr}
\end{figure*}

Finally, we also estimate the CBC \ac{sgwb} at multiple $\rho_{\rm thresh}$, which determines what we define as a resolvable signal. We note that our fiducial threshold of $\rho_{\rm thresh} = 12$ is somewhat conservative for the purposes of the unresolved \ac{sgwb}. Figure~\ref{fig:1CE40_1CE20_snr} shows the \textsc{pdb-pl} and the joint envelopes for two different SNR choices. The threshold $\rho_{\rm thresh} = 20$, in particular, was shown to be the frequency-independent optimal threshold for minimizing the effective background from BNS systems, including both the subtraction residue and the unresolved background~\cite{Zhou:2022nmt}.

\section{Implications and Conclusions} 
\label{sec:conclusion}

In this paper, we have presented a data-driven estimate of the strength of the unresolvable CBC background seen by XG detectors, accounting for both statistical and systematic uncertainties. A robust implication from our study is that $\Omega_{\rm cbc, unres}$ will be an impediment for the direct detection of many cosmological \ac{sgwb} models, independent of the fidelity and efficacy of the subtraction of resolved CBC signals. Some manner of simultaneous inference of the astrophysical and cosmological background (e.g., see Refs.~\cite{Martinovic:2020hru, Yu:2022xdw, Kaiser:2022cma}) will be required at a minimum.

Given the large number of resolvable CBC signals that we expect, it is also possible that their population inference can be used as a strong prior for $\Omega_{\rm cbc, unres}$. While this could make inferring multiple backgrounds easier, this effort could be hindered by the efficacy of subtraction. Moreover, using the resolvable signals to form a prior could create a systematic bias because the resolvable and unresolvable sources come from different redshifts~\cite{Callister:2016ewt}. 

However, this also makes $\Omega_{\text{cbc, unres}}$ astrophysically interesting in its own right by allowing us to study compact binary populations at high redshifts~\cite{Callister:2020arv, Mukherjee:2021itf}. Since we will confidently detect $\Omega_{\rm cbc, unres}$ with XG detectors, this represents a clear science case that can inform detector design. For this purpose, in Tab.~\ref{tab:fiducial_results} we provide figures-of-merit of the level of $\Omega_{\text{cbc, unres}}$ for various detector networks at two reference frequencies.

Our results also help inform the target efficacy of subtraction techniques for XG detectors. The optimal level of subtraction is one that leaves a residue not significantly higher than $\Omega_{\rm cbc, unres}$. At the same time, a subtraction technique that leaves a residue substantially lower than $\Omega_{\rm cbc, unres}$ will not be ideal and likely computationally expensive from a statistical point of view. Our figures-of-merit for the various networks in Tab.~\ref{tab:fiducial_results} will again be useful in informing the optimal efficacy of subtraction that algorithms should target.

\begin{table*}[ht]
    \centering
    \begin{tabular*}{0.95\textwidth}{c  @{\extracolsep{\fill}} c c c c } 
\hline \hline
\multicolumn{2}{c}{}  & \multicolumn{3}{c}{$\log_{10} \Omega_{\rm cbc, unres}$} \\ 
\cline{3-5} Model & $f$ & $\rho_{\rm thresh} = 12$ & $\rho_{\rm thresh} = 20$ & $\rho_{\rm thresh} = 40$ \\
\hline 
\multicolumn{1}{c}{}  & \multicolumn{3}{c}{40 km CE A, 20 km CE B, 10 km ET} \\
\hline \textsc{joint-envelope} & 25 Hz & $-10.8^{+0.4}_{-0.4}$& $-10.3^{+0.4}_{-0.4}$ & $-10.0^{+0.3}_{-0.4}$\\ 
& 250 Hz & $-10.3^{+0.5}_{-0.4}$& $-9.9^{+0.4}_{-0.4}$ & $-9.6^{+0.4}_{-0.4}$\\
\textsc{pdb-pl} & 25 Hz & $-10.4^{+0.4}_{-0.5}$& $-10.0^{+0.4}_{-0.4}$ & $-9.8^{+0.4}_{-0.4}$\\
 & 250 Hz & $-9.8^{+0.4}_{-0.5}$& $-9.5^{+0.4}_{-0.5}$ & $-9.3^{+0.4}_{-0.4}$ \\
\hline
\multicolumn{1}{c}{}  & \multicolumn{3}{c}{40 km CE A, 10 km ET, 4 km A\# LLO} \\  
\hline \textsc{joint-envelope} & 25 Hz & $-10.6^{+0.4}_{-0.4}$& $-10.3^{+0.4}_{-0.4}$ & $-10.0^{+0.4}_{-0.4}$\\ 
& 250 Hz & $-10.2^{+0.4}_{-0.4}$& $-9.8^{+0.4}_{-0.4}$ & $-9.6^{+0.4}_{-0.4}$\\
\textsc{pdb-pl} & 25 Hz & $-10.3^{+0.4}_{-0.5}$& $-10.0^{+0.4}_{-0.5}$ & $-9.8^{+0.4}_{-0.5}$\\ 
 & 250 Hz & $-9.8^{+0.4}_{-0.5}$& $-9.4^{+0.4}_{-0.5}$ & $-9.3^{+0.4}_{-0.5}$ \\ 
 \hline 
 \multicolumn{1}{c}{}  & \multicolumn{3}{c}{40 km CE A, 20 km CE B} \\
 \hline 
\textsc{joint-envelope} & 25 Hz & $-10.6^{+0.4}_{-0.4}$& $-10.2^{+0.4}_{-0.4}$ & $-10.0^{+0.4}_{-0.3}$\\
& 250 Hz & $-10.1^{+0.4}_{-0.4}$& $-9.8^{+0.4}_{-0.4}$ & $-9.6^{+0.4}_{-0.3}$\\ 
\textsc{pdb-pl} & 25 Hz & $-10.2^{+0.4}_{-0.5}$& $-9.9^{+0.4}_{-0.5}$ & $-9.8^{+0.4}_{-0.4}$\\
 & 250 Hz & $-9.7^{+0.4}_{-0.5}$& $-9.4^{+0.4}_{-0.5}$ & $-9.2^{+0.4}_{-0.4}$ \\ 
\hline  
\multicolumn{1}{c}{}  & \multicolumn{3}{c}{40 km CE A, 10 km ET} \\
\hline \textsc{joint-envelope} & 25 Hz & $-10.6^{+0.4}_{-0.4}$& $-10.3^{+0.4}_{-0.4}$ & $-10.0^{+0.4}_{-0.4}$\\
& 250 Hz & $-10.2^{+0.4}_{-0.4}$& $-9.8^{+0.4}_{-0.4}$ & $-9.6^{+0.4}_{-0.4}$\\ 
\textsc{pdb-pl} & 25 Hz & $-10.3^{+0.4}_{-0.5}$& $-10.0^{+0.4}_{-0.5}$ & $-9.8^{+0.4}_{-0.5}$\\
 & 250 Hz & $-9.8^{+0.4}_{-0.5}$& $-9.4^{+0.4}_{-0.5}$ & $-9.3^{+0.4}_{-0.5}$ \\
\hline  
\hline
\end{tabular*}
    \caption{$\Omega_{\text{cbc, unres}}$ for different networks at two reference frequencies for the \textsc{pdb-pl} model and the \textsc{joint-envelope}. The quoted numbers correspond to median values with $90\%$ uncertainties. We present these estimates as figures of merit for both detector design and subtraction algorithms.}
\label{tab:fiducial_results}
\end{table*}

Instead of subtraction, one can also infer the parameters of CBC signals without any threshold. This Bayesian ``global fit" technique was first developed in Ref.~\cite{Smith:2017vfk}, with various extensions studied in Refs.~\cite{Smith:2020lkj, Talbot:2021igi, Banagiri:2020kqd, Biscoveanu:2020gds}. Ref.~\cite{Biscoveanu:2020gds} in particular showed how a cosmological (non-CBC) background can be incorporated into this formalism for simultaneous inference. While this statistical formalism holds much promise because it removes both the need for subtracting the resolved CBC signals and for separately inferring the unresolved CBC background, it is also computationally very expensive, which can make running it on long stretches of data untenable. Moreover, the method is susceptible to subtle systematics~\cite{Talbot:2021igi} and can struggle to deal with overlapping signals, both of which can make an application on real XG data challenging.

Finally, we discuss some of the caveats of this study. One of the most important assumptions we made is the specific form of the SFR distribution. While we have tried two different SFR models as we show in Fig.~\ref{fig:1CE40_1CE20_sfr_td_mdr}, uncertainties on the rate naturally increase at higher redshift. For instance, observations with the James Webb Space Telescope show heightened star formation at high redshifts, which might imply a louder CBC background, and more observation in the near future might make the picture clearer~\cite{Finkelstein:2023, Harikane:2023}. Another effect that is somewhat related, which we did not consider here is the effect of metallicity~\cite{Lehoucq:2023zlt, Perigois:2020ymr, Kowalska-Leszczynska:2012gkb, Martinovic:2021fzj, Nakazato:2016nkj, Dvorkin:2016wac}. This would again predominantly impact the rate at higher redshifts. 

On the GW waveform side, we have assumed zero spins and tidal effects. While these are all subdominant compared to the mass distributions, it is possible that the spins in particular might have a noticeable impact on $\Omega_{\rm cbc, unres}$ given the lengths of the signals in XG detectors.

\section{Acknowledgements} 
\label{sec:acknowledgements}

We thank Salvatore Vitale, Emanuele Berti, Vuk Mandic, Haowen Zhong, and Vishal Baibhav for helpful discussions. We thank Amanda Farah and Philippe Landry for their help with the LVK population analyses and Sylvia Biscoveanu for her help with the NSBH population analyses and for providing PI curves for XG detectors. D.S.B. is supported by the National Science Foundation (NSF) Graduate Research Fellowship Program under grant DGE-2234667. Any opinion, findings, and conclusions or recommendations expressed in this material are those of the authors and do not necessarily reflect the views of the NSF. D.S.B. was supported by a CIERA Post-Baccalaureate Research Fellowship during the majority of this research. D.S.B. acknowledges partial support for this project from NASA through a NASA Illinois Space Grant. S.B. acknowledges support from the NSF grant PHY-2207945. Z.D. is supported by the CIERA Board of Visitors Research Professorship. V.K. is partially supported through a CIFAR Senior Fellowship, a Guggenheim Fellowship, the Gordon and Betty Moore Foundation (grant award GBMF8477), and from Northwestern University, including the Daniel I. Linzer Distinguished University Professorship fund. 

This material is based upon work supported by NSF's LIGO Laboratory which is a major facility fully funded by the National Science Foundation.  This research has made use of data obtained from the Gravitational Wave Open Science Center (\href{https://gwosc.org}{gwosc.org}), a service of LIGO Laboratory, the LIGO Scientific Collaboration, the Virgo Collaboration, and KAGRA. LIGO Laboratory and Advanced LIGO are funded by the United States NSF as well as the STFC of the United Kingdom, the Max-Planck-Society (MPS), and the State of Niedersachsen/Germany for support of the construction of Advanced LIGO and construction and operation of the GEO\,600 detector. Additional support for Advanced LIGO was provided by the Australian Research Council. Virgo is funded, through the European Gravitational Observatory (EGO), by the French Centre National de Recherche Scientifique (CNRS), the Italian Istituto Nazionale di Fisica Nucleare (INFN) and the Dutch Nikhef, with contributions by institutions from Belgium, Germany, Greece, Hungary, Ireland, Japan, Monaco, Poland, Portugal, Spain. KAGRA is supported by Ministry of Education, Culture, Sports, Science and Technology (MEXT), Japan Society for the Promotion of Science (JSPS) in Japan; National Research Foundation (NRF) and Ministry of Science and ICT (MSIT) in Korea; Academia Sinica (AS) and National Science and Technology Council (NSTC) in Taiwan. The authors are grateful for computational resources provided by the LIGO Laboratory and supported by NSF Grants PHY-0757058 and PHY-0823459. Lastly, we acknowledge the efforts of the \ac{ce} Consortium and \ac{et} Collaboration in the planning and development of XG GW detectors. This manuscript carries an LSC DCC number LIGO-P2300334.
\newpage
\bibliography{references}

\appendix

\section{Stochastic backgrounds from alternate detector networks}
\label{sec:appendix_A}

\begin{figure*}[hbtp]
    \centering
    \begin{subfigure}[b]{0.75\textwidth}
            \includegraphics[width=0.78\textwidth]{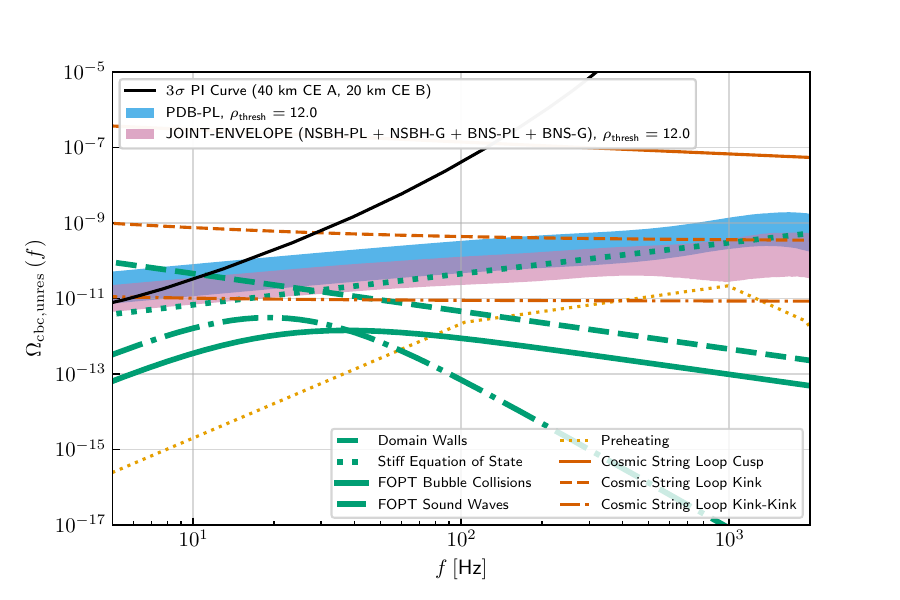}
            \caption{$\Omega_{\text{cbc, unres}}$ estimates for a network consisting of a 40-km CE A and a 20-km CE B.}
            \label{fig:2CE40_full_envelope}
    \end{subfigure}

    \begin{subfigure}[b]{0.75\textwidth}
            \includegraphics[width=0.78\textwidth]{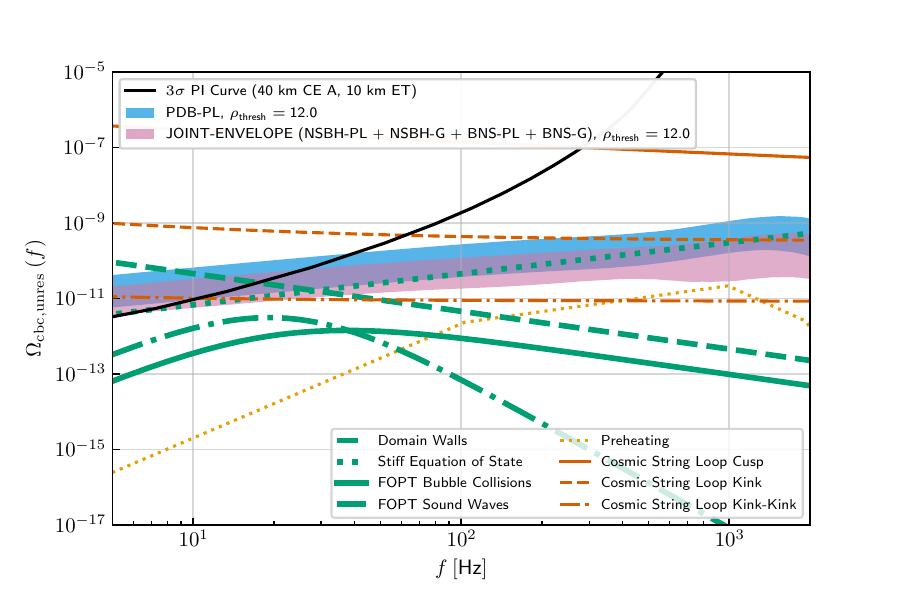}
            \caption{$\Omega_{\text{cbc, unres}}$ estimates for a network consisting of a 40-km CE A and an ET.}
            \label{fig:1CE40_1ET_full_envelope}
    \end{subfigure}

    \begin{subfigure}[b]{0.75\textwidth}
            \includegraphics[width=0.78\textwidth]{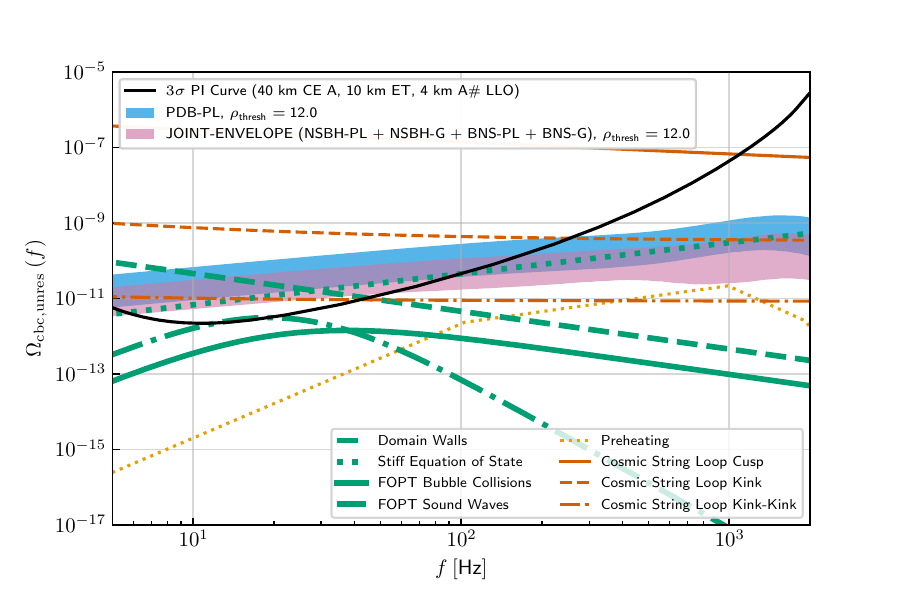}
            \caption{$\Omega_{\text{cbc, unres}}$ estimates for a network consisting of a 40-km CE A, an $\text{A}^{\#}$ at LLO, and an ET.}
            \label{fig:1CE40_1ASHARP_1ET_full_envelope}
    \end{subfigure}
        \caption{A comparison of various cosmological \ac{sgwb} models along with $\Omega_{\text{cbc, unres}}$ estimates for the various alternate detector networks we consider in Sec.~\ref{sec:detector_networks}. The blue band shows the \textsc{pdb-pl} model while the pink band shows the \textsc{joint-envelope} constructed from the various BNS-only and NSBH-only models in Tab.~\ref{tab:1}. Both bands represent $90\%$ credible intervals. The solid black curve is the $3 \sigma$ PI curve.}
    \label{fig:alternate_networks_full_envelope}
\end{figure*}

\begin{figure*}[ht]
    \centering
    \includegraphics[width=0.85\textwidth]{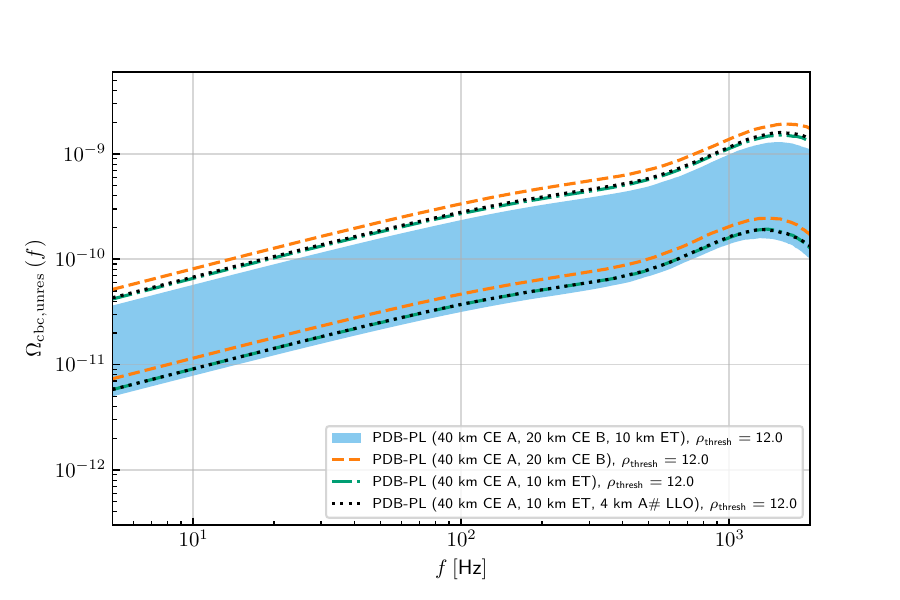}
    \caption{A comparision of the 90\% credible intervals of $\Omega_{\text{cbc, unres}}$ for the \textsc{pdb-pl} model for the four different detector networks defined in Sec.~\ref{sec:detector_networks}.}
    \label{fig:pdb_network_comparison}
\end{figure*}

In this appendix, we show the results for the alternate detector networks that we consider in this study. Figure~\ref{fig:alternate_networks_full_envelope} shows our $\Omega_{\text{cbc, unres}}$ estimates for the \textsc{pdb-pl} and \textsc{joint-envelope} models for the various alternate detector networks we consider in Sec.~\ref{sec:detector_networks}. Also overlaid are the different cosmological models discussed in Sec.~\ref{sec:cosmo_backgrounds}. The equivalent plot for our fiducial network is Fig.~\ref{fig:fiducial_joint_unresolved}. Finally, Fig.~\ref{fig:pdb_network_comparison} overlays the \textsc{pdb-pl} model $\Omega_{\text{cbc, unres}}$ estimates for all our considered networks as a means of direct comparison.

\end{document}